# Ultrafast Proton Transport between a Hydroxy Acid and a Nitrogen Base along Solvent Bridges Governed by Hydroxide/Methoxide Transfer Mechanism

Maria Ekimova[1], Felix Hoffmann[2], Gül Bekçioğlu-Neff[2], Aidan Rafferty[1], Oleg Kornilov[1], Erik T. J. Nibbering[1], Daniel Sebastiani[2]

[1]*Max Born Institut für Nichtlineare Optik und Kurzzeitspektroskopie, Max Born Str. 2A, 12489 Berlin, Germany*
[2]*Institut für Chemie, Martin-Luther-Universität Halle-Wittenberg, Von-Danckelmann-Platz 4, 06120 Halle (Saale), Germany*
E-mail: nibberin@mbi-berlin.de ; daniel.sebastiani@chemie.uni-halle.de

Current text is 3028 words

Number of Figures + Tables is now 6  (5 Figures + 1 Table)

**Abstract:**
Aqueous proton transport plays a key role in acid-base neutralization, and energy transport through biological membranes and hydrogen fuel cells. Extensive experimental and theoretical studies have resulted in a highly detailed elucidation of one of the underlying microscopic mechanisms for aqueous excess proton transport, known as the von Grotthuss mechanism, involving different hydrated proton configurations with associated high fluxional structural dynamics. Hydroxide transport, with approximately two-fold lower bulk diffusion rates than those of excess protons, has received much less attention. We present femtosecond UV/IR pump-probe experiments and *ab initio* molecular dynamics simulations of different proton transport pathways of the bifunctional photoacid 7-hydroxyquinoline (7HQ) in water-methanol mixtures. For 7HQ solvent-dependent photoacidity, free energy-reactivity correlation behaviour and QM/MM trajectories point to a dominant $OH^-/CH_3O^-$ transport pathway, for all water-methanol mixing ratios investigated. This provides conclusive evidence for the hydrolysis/methanolysis acid-base neutralization pathway formulated by Manfred Eigen half a century ago.

## Main

Aqueous acid-base neutralization involves a proton exchange with strong involvement of water solvent molecules. The current understanding of proton exchange between Brønsted acids and Brønsted bases originated from seminal studies by Eigen [1] and Weller [2] more than half a century ago. The general kinetic scheme has been described in Eigen's review of the field [3]. According to the scheme, the proton transfer converting the acid-base reactant pair into its conjugate acid-conjugate base product pair is considered to follow either concerted or sequential pathways (**Figure 1a**). In the sequential (proton hopping) case two possible directions for proton transport can in principle occur: either a forward excess proton transfer from acid via intermediate hydrated protons to the accepting base (protolysis pathway), or an inverse proton defect transport (hydrolysis pathway) where the base extracts a proton from the solvent, generating hydrated hydroxide anions, that ultimately react with the acid [4,5]. A third pathway involves a direct proton exchange between acid and base, with an underlying concerted proton transfer mechanism. Which proton transfer pathway prevails depends on relative reaction rates of the individual proton transfer steps. The relevance of proton transport in cases as diverse as hydrogen fuel cells and biochemical environments, including transmembrane protein channels [6-9], necessitates a further exploration of the underlying proton transport mechanisms for conditions clearly distinct from those of bulk water, i.e., for lower polarity solvent media and possible crucial roles of hydrophobic alkyl- and hydrophilic alcohol-functionalities.

Experimental and theoretical results point to a water-mediated proton transfer mechanism on ultrafast time scales between aromatic alcohols and carboxylate bases [10-15]. A limited number of intermediate water molecules connecting these acid and base molecules facilitate the protolysis pathway through a hydrogen-bonded "water bridge" or "water wire". These findings can be compared to theoretical studies on proton transport in bulk water, where the transport mechanism of the excess proton has been intensively explored [16-18]. It has been found that different conformations of hydrated protons play a key role in the sequential von Grotthuss-type proton hopping mechanism, with Eigen-type (hydronium, $H_3O^+$) or Zundel-type ($H_5O_2^+$) hydrated protons often found to occur at distinct phases of the elementary proton hopping events [19-22]. Hydroxide ($OH^-$) transport has only been considered in a small number of *ab initio* molecular dynamics studies [23-25], where for $OH^-$ transport in bulk water two different hydrated forms of $OH^-$ have been found. These studies found $OH^-$-transfer to occur by either 4- or 5-fold hydrated configurations, with $OH^-$ donating one and accepting 3 or 4 hydrogen bonds to first solvation shell water molecules, respectively.

In this report we present results of a joint experimental and theoretical study of proton transfer between a proton donating OH-group of an acid and a proton accepting aromatic N-moiety of a base. To avoid a possible variation of the distance between the acid and base, thus maintaining a well-defined number of solvent molecules forming the solvent bridge, we use a so-called bifunctional photoacid [26], 7-hydroxyquinoline (7HQ) [27-32] (**Figure 1b**). This bifunctional compound has both the photoacid properties of 2-naphthol and the photobase properties of quinoline. Excitation to the first electronically excited state results in a p$K_a$-jump from 8.67 to around 0.4 for the OH-group and from 5.64 to around 11.1 for the N side (see **Table 1**). As a result, both the acidity of the OH-group and the basicity of the N-group strongly increase, initiating a net proton transfer from the neutral (N*) to the zwitterionic (Z*) form with a time constant of ~ 37 ps in water [28] and 170 ps in methanol [29]. In a previous study we have characterized 7HQ dissolved in deuterated methanol ($CD_3OD$) [33], and found the IR-active fingerprint patterns of N*, Z*, and intermediates A* or C* that are expected to occur when the reaction proceeds through a forward or an inverse proton transfer pathway, respectively (**Figure 1c**), where, however, the results have not been sufficient to draw definite conclusions on the dominant reaction pathway.

In this work we follow the conversion of N* → Z* in real time for different water-methanol mixtures, further exploring the N* → Z* conversion kinetics with ultrafast UV/IR pump-probe spectroscopy. We characterize the reaction rates as a function of $X_{H_2O}:X_{CH_3OH}$ molar fraction ratio, and look for possible presence of transient species. To substantiate the observed reaction rates we derive the differences in acidity of reactants and products and possible intermediates through the respective p$K_a$/p$K_a$*-values using well-established free energy-reactivity correlations [12, 34-37]. We conclude that for ensemble averaged population kinetics upon photoexcitation of 7HQ, the proton transfer follows general rules for acidities when going from methanol to water. To further substantiate our findings that the proton transfer pathway through a sequential hydroxide/methoxide transport occurs for all $X_{H_2O}:X_{CH_3OH}$ molar fraction ratios we use QM/MM molecular dynamics simulations applied to the $S_1$-state of 7HQ using time-dependent density functional theory (TD-DFT) calculations. We infer that the primary events of proton transfer for 7HQ involves a proton abstraction from the nearest solvent molecule to the quinoline N-side. Our combined experimental and theoretical results demonstrate that the hydrolysis/methanolysis pathway of acid-base neutralization through a water/methanol bridge consisting of three solvent molecules is the dominant pathway. Our results on the 7HQ model system have not only direct relevance for proton transport in water-rich solutions, but also for less polar water poor media, that are ubiquitous in the important cases of proton transport channels in transmembrane proteins or ion exchange regions within hydrogen fuel cells.

## Results and Discussion

We have measured the transient response of the vibrational marker modes of 7HQ in the IR-active fingerprint spectral region between 1400 and 1550 cm$^{-1}$. In a previous publication [33] we have reported on the distinct vibrational patterns for the different charged species N*, C*, A* and Z*, dissolved in neat CD$_3$OD. We have used deuterated methanol as a solvent, to facilitate a direct access to the most important vibrational marker modes in this spectral region, which for normal methanol CH$_3$OH would be inaccessible under our experimental conditions. We have concluded that the deuteron transfer process occurs from N* → Z* with a 330 ps time constant. **Figure 2a** shows our transient IR-absorption spectra, with basic features similar for all water-methanol mixtures studied, i.e. for the $X_{D_2O}:X_{CD_3OD}$ molar ratio ranging from 0.0:1.0 to 0.7:0.3. Indeed, we identify the N* marker mode located at 1475 cm$^{-1}$, and the Z* marker mode at 1440 cm$^{-1}$, to let us observe a profound increase in overall conversion rates from N* → Z* with increasing water content (**Figure 2b**). For all water-methanol solvent mixtures the time scale of the decay of N* population is equal to the rise of Z* population. We have not observed significant spectral features indicative of transient population build-up of either the C* or A* species upon increased molar fraction of water. From these results we conclude that the underlying proton transport mechanisms for 7HQ in water-methanol mixtures must be similar to those in neat methanol, up to the highest water molar fraction investigated here. An in-depth kinetic analysis of the transient IR spectra can be found in the SI. We summarize our findings of this kinetic analysis in **Table S1**, and compare the obtained time constant values with those previously reported.

7HQ in principle can exhibit acid-base reactivity as an aromatic alcohol (through its 2-naphthol functionality) and as a protonated nitrogen base (through its quinolinium functionality). Free energy-reactivity correlations connect acid-base reaction rates with p$K_a$-values [38, 39]. Empirical data have been gathered on the solvent dependence of the reactivity of acids and bases [40-42]. A determination of p$K_a$*-values of 7HQ as a function of solvent mixture composition should thus reveal which reaction pathway (proton transport via pathway I, or inverse proton transport via pathway II) dominates. We use free energy-reactivity correlation curves – well-established in photoacid research [12, 34-39, 43, 44] – to determine the overall p$K_a$*-values governing the proton transfer dynamics in water, in methanol, and in water-methanol mixtures. For details we refer to the Supplemental Information. **Figure 3** depicts our findings on 7HQ dissolved in neat CD$_3$OD as well as the $X_{D_2O}$:$X_{CD_3OD}$ solvent mixtures using our femtosecond UV/IR pump-probe measurements, for the overall N* ⇆ Z* proton transfer reaction, as well as for the C* ⇆ Z* and A* ⇆ Z* steps in neat CD$_3$OD. For comparison we also have added previously reported results by other research groups obtained with time-correlated single-photon counting (TCSPC) measurements [28, 29]. It follows that the measurements performed in H$_2$O, in CH$_3$OH and in CD$_3$OD all follow the same free energy-reactivity correlation, provided that the N* ⇆ Z* reaction underlies a change in value of Δp$K_a$ = p$K_a$(donor) – p$K_a$(protonated acceptor), by increasing about 1.2 – 1.3 units when going from H$_2$O to CH$_3$OH (or, equivalently, from D$_2$O to CD$_3$OD) as solvent. This result strongly points to the quinoline unit dominating the overall proton transfer dynamics for the N* → Z* conversion, as for the naphthol unit a shift of 3 – 5 units is expected [40-42]. Hence we conclude that our experimental results are indicative of a proton transport mechanism via the N* → C* → Z* pathway II, i.e. an inverse proton transfer mechanism by hydroxide/methoxide transfer. An additional increase in magnitude of 0.3 – 0.5 units for Δp$K_a$ = p$K_a$(donor) – p$K_a$(protonated acceptor) occurs when one changes CH$_3$OH with CD$_3$OD [34]. As we do not observe any irregularities in the free energy-reactivity correlation for 7HQ in the water-methanol mixtures, we derive that at the level of our ensemble-averaged observations of the proton/deuteron transfer reaction dynamics, no preference can be concluded for proton transfer via either water or methanol.

To investigate further the underlying microscopic mechanisms of proton transport between donating and accepting groups of 7HQ in water-methanol mixtures, we have performed equilibrium molecular dynamics (MD) simulations to investigate the microscopic solvation of 7HQ, adiabatic MD calculations using a TD-DFT molecular mechanics (MM) approach to identify the propensity of particular ultrafast protonation events of 7HQ in the S$_1$-state, and AIMD simulations to obtain key insight into the proton transport along a solvent bridge, following the initial proton transfer between 7HQ and the solvent.

Microsolvation around 7HQ, in particular along the solvent bridge connecting the OH and N moieties, appears to show a bias to a larger number of water molecules than the statistical value given the $X_{H_2O}$:$X_{CH_3OH}$ molar fractions used, even though the actual numbers depend on using either ab initio MD or classical MD routines to determine this. For the $X_{H_2O}$:$X_{CH_3OH}$ mixtures investigated here, 0.5:0.5 and 0.7:0.3, the dominant configurations involve two or three water molecules, with ones consistent of methanol only playing a minor role. From our simulations we deduce that solvent bridges with full hydrogen bonds ("solvent wires") are only a minor part (7 – 17 %) for the $X_{H_2O}$:$X_{CH_3OH}$ = 0.5:0.5 and 0.7:0.3 molar fraction ratios. Hence the most typical solvent bridge configurations involve two to three water molecules with partially broken hydrogen bonds. Further details are provided in the SI.

To investigate the mechanism of N* to Z* conversion, we carried out first principles adiabatic MD simulations of the fully solvated 7HQ in the first electronically excited state, employing TD-DFT. The main focus was on whether the reaction proceeds concertedly via a solvent wire or sequentially in a von-Grotthuss-type hopping mechanism. In the case of a sequential reaction, there are two further possibilities, namely $H_3O^+/MeOH_2^+$ or $OH^-/CH_3O^-$ transport. For this reason, we prepared the system in two initial conditions that were based on AIMD snapshots of N in the ground state: (I.) C* with a hydrogen bonded $OH^-/CH_3O^-$ at the nitrogen site, and (II.) A* with an $H_3O^+$ molecule hydrogen bonded to the A*-$O^-$ site (cf. **Figure 4**). Moreover, to allow for the possibility of a concerted reaction, all of the selected initial configurations exhibited a solvent wire which connected the photobasic and photoacidic sites of 7HQ. The system was propagated adiabatically on the excited state potential energy surface for the duration of about 1 ps, whilst the migration of the excess charge was followed. This approach allows us to assess the temporal stability of C* and A* protonation states inside a realistic solvent environment and helps to address the question which of the possible reaction pathways is the most likely.

The simulations reveal that for most trajectories with C* as the initial state, no back transfer of the proton to the adjacent negatively charged solvent ion was observed. In contrast, the proton was transferred back to the basic 7HQ oxygen atom already within the first 200 fs for trajectories if the initial state was A*. For one of the A* trajectories, we observed the transient formation of C* which was formed after back-protonation of the 7HQ hydroxyl group. Hence, an anionic charge transfer through the solvent, i.e. based on $OH^-$ and $CH_3O^-$ ions, is favoured, whereas the cationic variant, featuring $H_3O^+$ or $CH_3OH_2^+$ species, is clearly disfavoured. Although our MD simulations yield a consistent picture with our experimental results, we emphasize here that significantly longer MD simulations are necessary to unequivocally assess whether there exists a pathway for a concerted transfer, or not.

To follow the picosecond dynamics of the negative solvent ion ($OH^-/CH_3O^-$) beyond the ultrafast time scales accessible by the TD-DFT MD simulations, we carried out further AIMD simulations employing the revPBE0 hybrid functional and D3 dispersion corrections. To elucidate the underlying transfer mechanism, the hydrogen bonding configurations of the reacting species at various times during the transfer were analyzed. The left panel of **Figure 5** shows the average number of hydrogen bonds for the proton-donating solvent molecule (**Figure 5a**) and the solvent ion (**Figure 5b**), respectively. Note, it was not distinguished whether the solvent ion was methoxide or hydroxide. It can be seen that due to the lower density in the case of the $X_{H_2O}:X_{CH_3OH}$ = 0.5:0.5 mixture, the number of accepting hydrogen bonds is reduced by about 0.2. However, when the proton is symmetrically shared between the two solvent molecules, i.e. for structures close to the transition state, the number of hydrogen bonds is the same for both water-methanol ratios. Therefore, the simulations strongly support that the same underlying transfer mechanism exists in both cases. During the transfer, the number of accepted hydrogen bonds at the donor molecule increases to its ideal number 2. On the other hand, the number of accepting hydrogen bonds at the solvent ion drops on average to 2.9 at the transition state. This has important consequences. In the case of the $X_{H_2O}:X_{CH_3OH}$ = 0.5:0.5 solution, larger changes of the hydrogen bonding network are required at the donor molecule because the number of accepting hydrogen bonds is less due to the lower density. The situation is reversed for the $X_{H_2O}:X_{CH_3OH}$ = 0.7:0.3 mixtures. Here, the changes of the hydrogen bonding configuration are most pronounced at the solvent ion. This is because at higher water concentrations there is an increasing number of hyper-coordinated methanolate or hydroxide ions, i.e. structures that accept 4 hydrogen bonds. Consequently, the coordination number difference between solvent ions in their minimum free energy and their transition state configuration also increases. Therefore, our simulations reveal that two mechanistically counteracting effects are occurring when altering the water-methanol ratio. Notably, the changes in hydrogen bonding are larger in the case of the $X_{H_2O}:X_{CH_3OH}$ = 0.5:0.5 solutions that may be one reason for the lower conductivities in this case.

Our results are in line with what one would expect from the presolvation concept proposed by Tuckerman *et al.* [23, 24]. However, these studies pointed out that in the dynamical hypercoordination mechanism of hydroxide ion migration in water a donating hydrogen bond of the hydroxide ion plays an important role. To investigate this for our water-methanol mixtures, we computed the average number of donating hydrogen bonds of the hydroxide ion for different stages of the reaction. Note, methoxide lacks the possibility to donate a hydrogen bond. The results are shown in the right panel of **Figure 5**. It can be seen that with increasing water concentration the average number of hydrogen bonds increases drastically. The relative frequency more than doubles when going from $X_{H_2O}:X_{CH_3OH}$ = 0.5:0.5 to 0.7:0.3 molar fractions. Moreover, it increases significantly for structures that share the proton symmetrically (low δ-values), but the relative frequencies of a donating hydrogen bond are still low for both ratios. Hence, the donating hydrogen bond of the hydroxide ion only plays a minor role in the case of the investigated concentration ratios, and, therefore, is not a necessary condition for the charge transfer in water-methanol mixtures at the concentrations investigated here.

## Conclusions

We have shown that for proton exchange between the proton-donating naphthol OH-group and the proton-accepting quinoline N-moiety of the bifunctional photoacid 7HQ, linked together by a water/methanol solvent bridge, occurs by a sequential hydroxide/methoxide transport mechanism taking place on ensemble-averaged time scales of tens to hundreds of picoseconds. The observation that charge migration along a preformed solvent bridge at the hydrophobic 7HQ-solvent interface proceeds in a step-wise manner has direct implications for the understanding of charge migration along other hydrophobic interfaces, such as in proton channels, where there is still an ongoing debate whether concerted proton transfer is the reason for the high efficiencies of biological proton channels. Free energy-reactivity correlations show that for all $X_{H_2O}:X_{CH_3OH}$ mixing ratios bulk acidities (p$K_a$-values) follow general trends in solvent polarities, strongly advocating for further studies of the applicability of such correlations between acidities and reaction time scales for proton transport near apolar regions, as well as polar or ionic functionalities within transmembrane proteins and hydrogen fuel cells. Finally our combined experimental and theoretical study shows the equal importance of the hydrolysis/methanolysis as possible alternative pathway to the more intensively studied protolysis pathway for acid-base neutralization in protic solvents as originally formulated by Eigen.

## Supplementary Information

Experimental Details; Principal Component Analysis of 7HQ UV/IR Pump-Probe Data; Free Energy-Reactivity Analysis; Computational Details; Microsolvation around 7HQ; Supplementary Figures S1-S9; Supplementary Tables S1-S4; Supplementary References S1-S56.

# Acknowledgement

This work has benefitted from financial support by the German Science Foundation (project no. NI 492/13-1/SE 1008/11-1) and the European Research Council (ERC) under the European Union's Horizon 2020 research and innovation programme (ERC grant agreement no. 788704; E.T.J.N.). F.H. cordially thanks the Fonds der Chemischen Industrie for a Kekulé fellowship. A.R. cordially thanks the European Commission for an Erasmus+ Trainee Fellowship. We are grateful for stimulating discussions with Prof. Dr. Ehud Pines and Dr. Dina Pines. E.T.J.N. dedicates this work to the memory of Prof. Dr. Dan Huppert, pioneer of ultrafast photoacid spectroscopy.

**Table 1** Comparison of previously reported time constants obtained from ultrafast spectroscopic measurements on 7HQ, and derived values for $\Delta pK_a = pK_a$(donor) – $pK_a$(protonated acceptor).

| Equilibrium | Functionality | CD$_3$OD | | | CH$_3$OH | | | H$_2$O | | |
|---|---|---|---|---|---|---|---|---|---|---|
| | | Time constant $\tau$ (ps) [a] | -log$_{10}$[$k_r$] | $\Delta pK_a$ | Time constant $\tau$ (ps) [b] | -log$_{10}$[$k_r$] | $\Delta pK_a$ | Time constant $\tau$ (ps) [c] | -log$_{10}$[$k_r$] | $\Delta pK_a$ |
| C* + ROH ⇆ Z* + ROH$_2^+$ | 2-naphthol | 160 | 9.80 | 1.0 | 114 [d] | | | 18 | 10.74 | -1.0 |
| N* + ROH ⇆ C* + RO$^-$ | 1-quinolinium | 320 | 9.49 | 1.4 | 170 | 9.77 | 1.0 | 37 | 10.43 | -0.25 |
| N* + ROH ⇆ A* + ROH$_2^+$ | 2-naphthol | | | | | | | | | |
| A* + ROH ⇆ Z* + RO$^-$ | 1-quinolinium | 600 | 9.22 | 1.75 | 428 [d] | | | 180 | 9.74 | 1.0 |

[a]: Ref. 33 from femtosecond UV/IR measurements; [b]: Ref. 29 from time-correlated single photon counting (TCSPC) TCSPC measurements; [c]: Ref. 28 from TCSPC measurements; [d]: assuming regular H/D kinetic isotope effect.

Figure 1: (a) Eigen's reaction scheme ; (b) Eigen's scheme adapted to 7HQ ; (c) forward excess proton transfer vs. inverse proton defect transfer with sequence of proton transfer events along solvent bridge for 7HQ

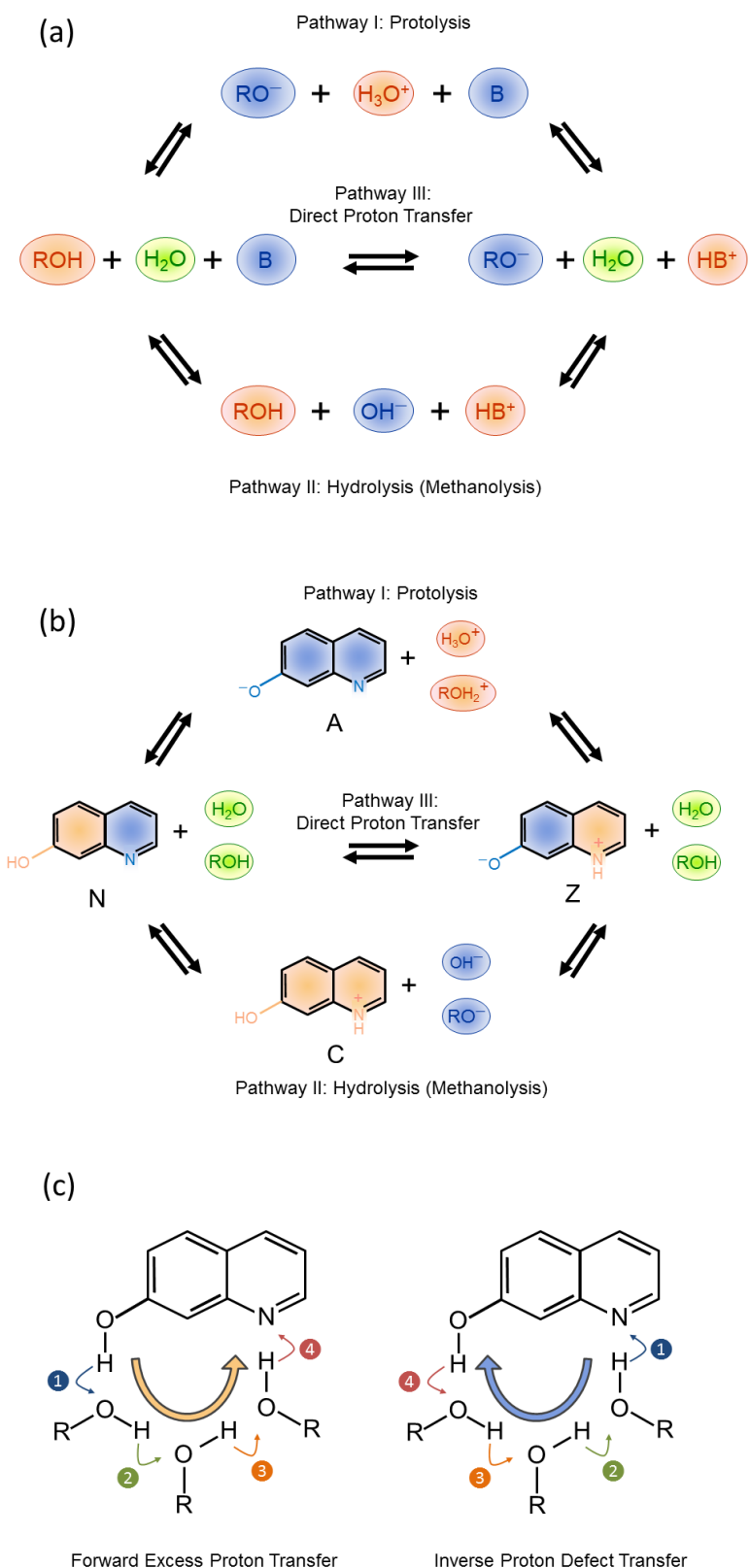

Figure 2: (a) Transient UV/IR spectra measured on 7HQ dissolved in D$_2$O-CD$_3$OD mixtures. (b) Population kinetics of the N* and Z* species, as measured through the 1475 cm$^{-1}$ and 1440 cm$^{-1}$ marker bands, respectively.

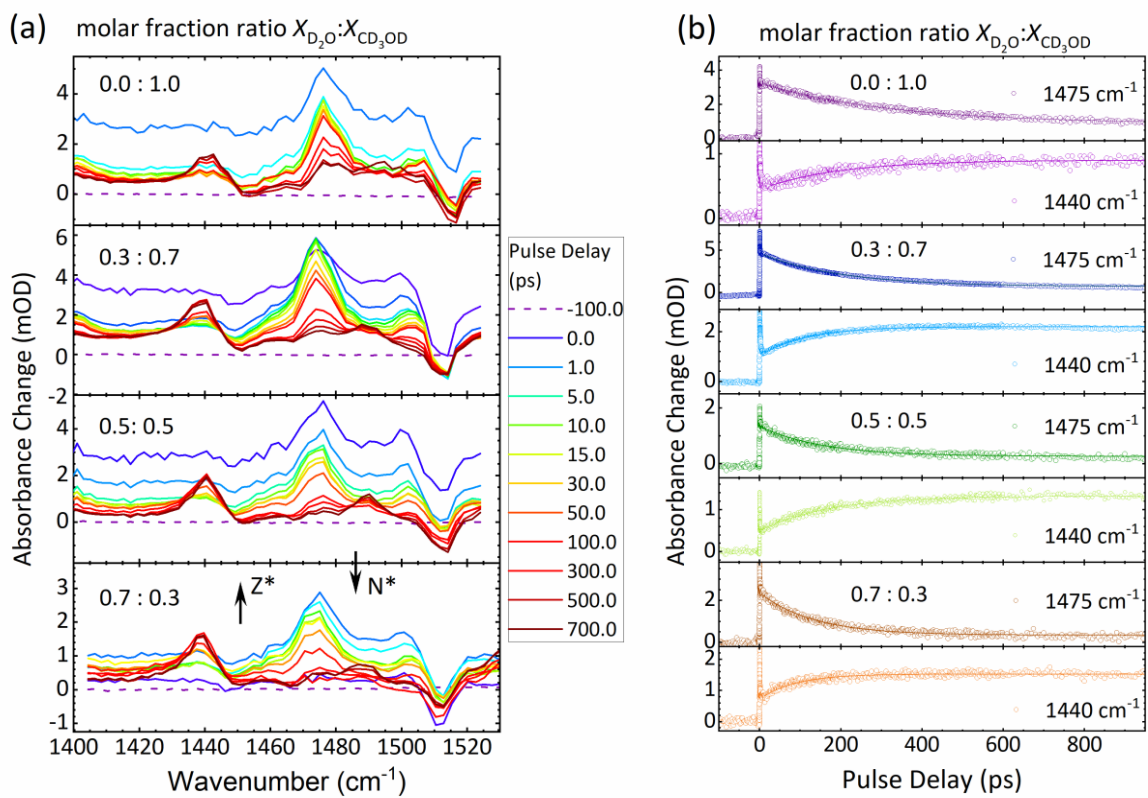

Figure 3: Free energy-reactivity correlation of the different charged forms of photoexcited 7HQ, in $H_2O$, $CH_3OH$ or $CD_3OD$ (a) and for $D_2O$-$CD_3OD$ mixtures (b). The Marcus Bond Energy Bond Order (BEBO) relationship is shown as solid line.

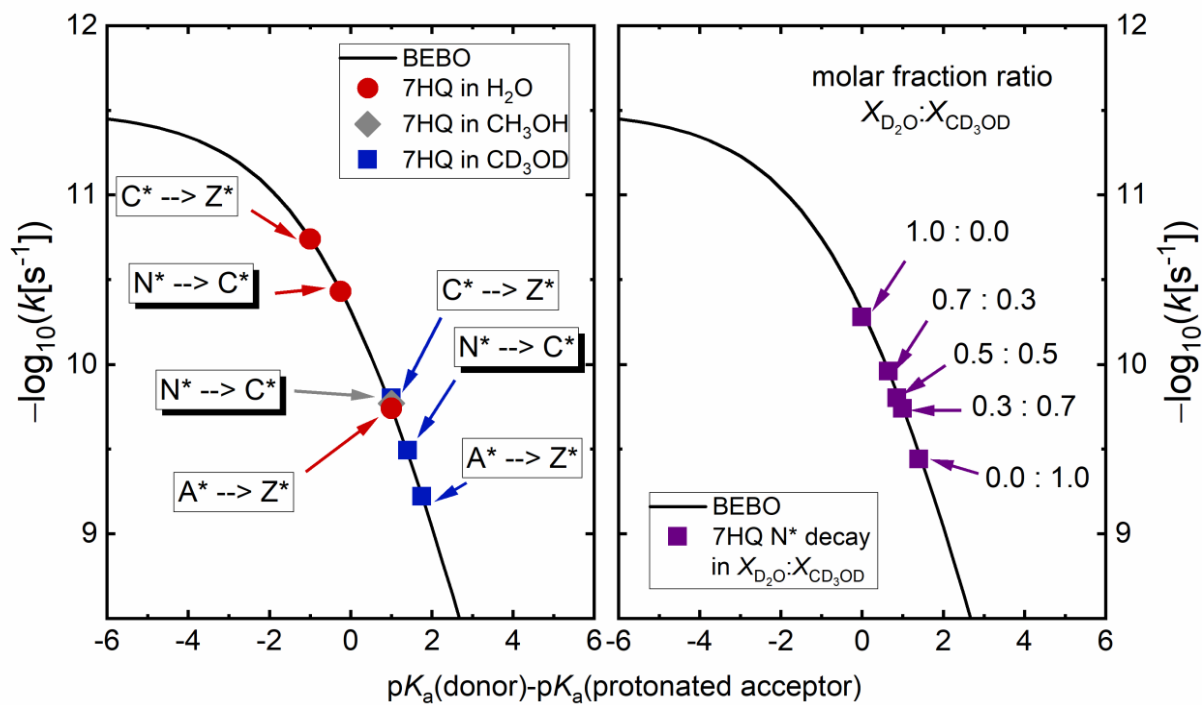

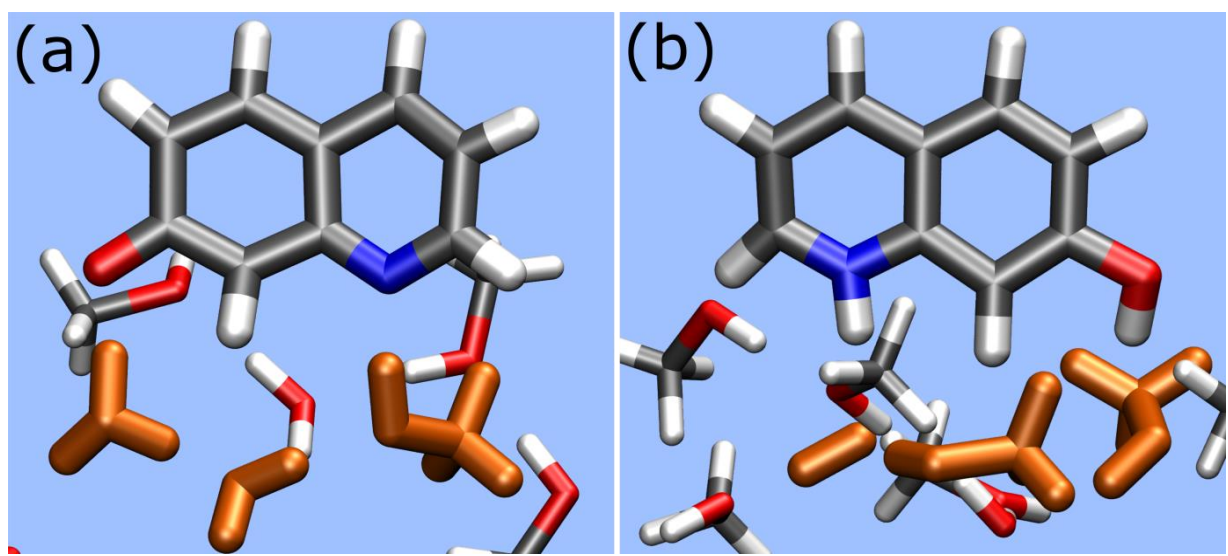

**Figure 4**  TD-DFT MD snapshots of A* (left) and C* (right) with surrounding solvent molecules. Molecules forming the wire are shown in orange. Only a limited number of QM atoms are shown for simplicity.

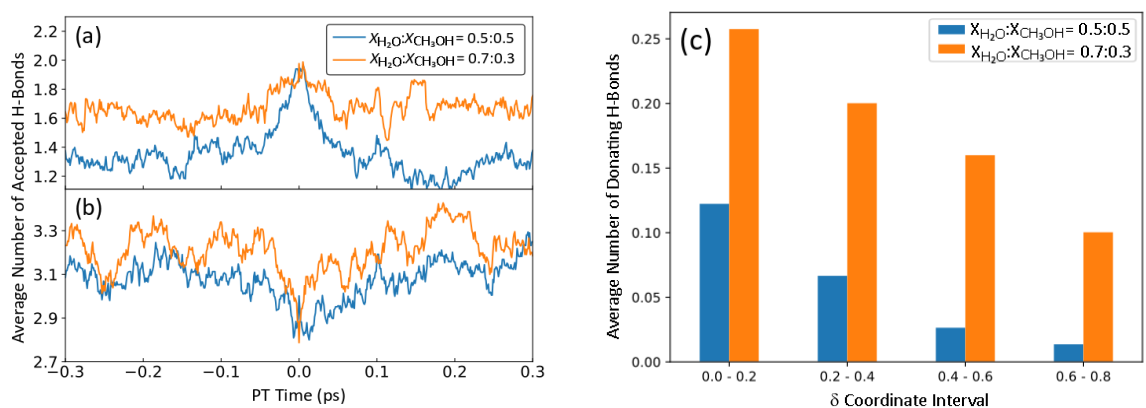

Figure 5 Average number of accepted hydrogen bonds of the proton donating solvent molecule (a) and the solvent ion ($OH^-$ or $CH_3O^-$) (b) as a function of the proton transfer time. The time origin is the time when the proton is symmetrically shared between donating and accepting oxygen atoms. Note that the proton donating solvent molecule and the solvent ion definitions interchange their respective roles for times > 0. (c) Average number of donating hydrogen bonds of the $OH^-$ ion as a function of the asymmetry parameter $\delta$ (for definition see SI) for the two mixing ratios. A small $\delta$-value indicates structures close to the transition state, where the proton is symmetrically shared between the two oxygen atoms, whereas larger values indicate structures close to free energy minima.

**Supplemental Information to:**

**Ultrafast Proton Transport between a Hydroxy Acid and a Nitrogen Base along Solvent Bridges Governed by Hydroxide/Methoxide Transfer Mechanism**

**Maria Ekimova[1], Felix Hoffmann[2], Gül Bekçioğlu-Neff[2], Aidan Rafferty[1], Oleg Kornilov[1], Erik T. J. Nibbering[1], Daniel Sebastiani[2]**

[1]*Max Born Institut für Nichtlineare Optik und Kurzzeitspektroskopie, Max Born Str. 2A, 12489 Berlin, Germany*
[2]*Institut für Chemie, Martin-Luther-Universität Halle-Wittenberg, Von-Danckelmann-Platz 4, 06120 Halle (Saale), Germany*
*E-mail: nibberin@mbi-berlin.de ; daniel.sebastiani@chemie.uni-halle.de*

Contents:

**1. Experimental Details**
**2. Principal Component Analysis of 7HQ UV/IR Pump-Probe Data**
**3. Free Energy-Reactivity Analysis**
**4. Computational Details**
**5. Microsolvation around 7HQ**

# 1. Experimental Details

7-hydroxyquinoline (99%) was purchased from ACROS Organics; $CD_3OD$ (99.8 %) and $D_2O$ (99.95 %) were purchased from Deutero and used without further purification. The typical concentration of the dye for the pump-probe experiments was 40-50 mM.

Steady-state absorption and fluorescence spectra were recorded on a Perkin Elmer spectrometer and a JOBIN YVON Horiba Fluorolog, respectively. The concentration of samples was $10^{-5}$ M guaranteeing an optical density below 0.5 at the band maximum in a 1 cm quartz cuvette.

In Figure S1 we show the absorption and fluorescence spectra of 7HQ in water/methanol mixtures. The gradual shift in equilibrium constants upon increased water fraction leads to a pronounced change in relative population of N and Z, and N* and Z*, in the $S_0$- and $S_1$-states, respectively.

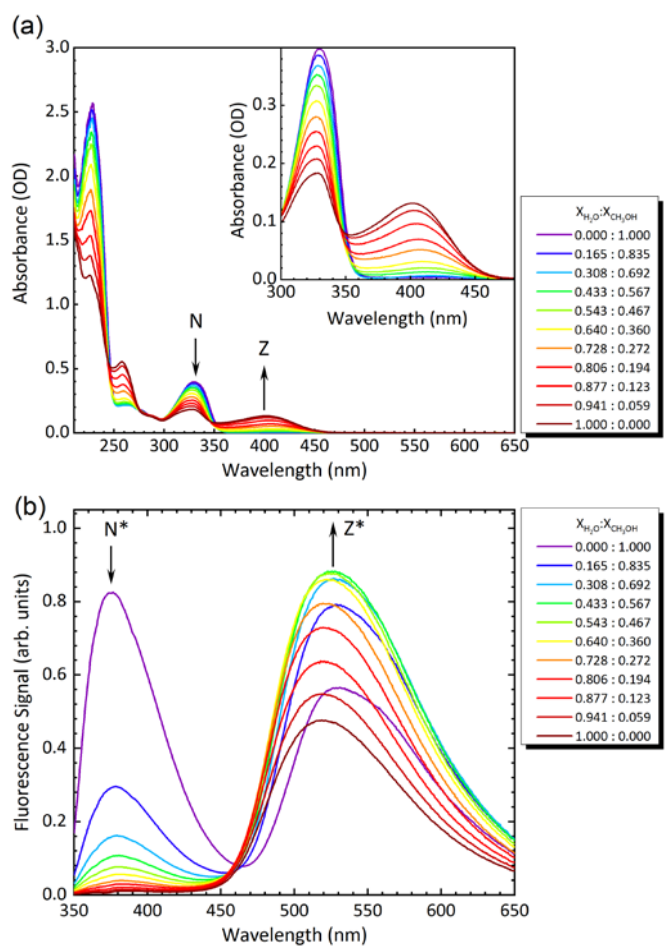

Figure S1: Absorption (a) and emission (b) spectra recorded for 7HQ in water-methanol mixtures with molar fractions $X_{H_2O}:X_{CH_3OH}$ as indicated in the Figure.

Ultrafast UV-pump IR-probe measurements were performed as described previously [1]. Briefly, the 340 nm pump beam was generated from an optical parametric amplifier (TOPAS-C , Light Conversion), pumped by the 800 nm beam at the repetition rate of 1 kHz (Tsunami oscillator with Spitfire Pro regenerative and booster amplifier stages, Spectra Physics), to have approximately 1 µJ at the sample. The pump pulses were stretched by a cuvette with water, resulting in a pump-probe cross correlation width of about 350 fs. The pump beam was temporally delayed with respect to the probe beam by a delay stage and then focused onto the sample. Tunable IR pulses were generated by a home-built double-pass optical parametric amplifier, followed by difference frequency mixing of signal and idler. The output beam was sent to a ZnSe wedge and the reflected pulses from its front and back surfaces were used as probe and references beams. Both probe and reference pulses were focused onto the sample by means of an off-axis parabolic mirror (focal diameter about 200 µm) and after passing the sample were dispersed in a polychromator (2 cm$^{-1}$ resolution in the 1200-1700 cm$^{-1}$ frequency window). Spectrally resolved absorbance changes were recorded using a liquid nitrogen cooled HgCdTe double array detector (2 × 64 pixels).

To avoid any possible sample damage during the pump-probe measurement, the solution was constantly flowing in the CaF$_2$ windowed cell of 50 um thickness by means of a peristaltic pump. FT-IR spectra were measured before and after each pump-probe measurement to ensure that there was no sample structure change upon excitation or over time.

## 2. Principal Component Analysis of 7HQ UV/IR Pump-Probe Data

One of the most reliable ways to extract detailed kinetics information from transient spectrally-resolved pump-probe measurements is to carry out principal component analysis (PCA) of the experimental data [2,3]. This method relies on the so-called "bi-linear assumption", which requires that the experimental data can be represented as a sum of (possibly overlapping) transients with time-independent spectral fingerprints. Mathematically PCA relies on the method of singular value decomposition (SVD), which represents the data matrix M as a product:

$$M = U \times S \times V \qquad , \qquad (S1)$$

where columns of matrix U are the spectral components, the rows of matrix V are the corresponding transients and S is the diagonal matrix of singular values, which give weights to the transient maps in the data matrix. Decomposition of the matrix into the SVD form can immediately be used to decide on the number of species contributing to the observed kinetics. The number of species corresponds to the number of significant singular values in the matrix S. Note, that this method is not suitable if frequency shifts or spectral reshaping (due to e.g. spectral diffusion due to solvation phenomena, intramolecular vibrational redistribution (IVR) and/or vibrational cooling due to solute-solvent vibrational energy exchange) is observed, since such phenomena do not conform to the bi-linear assumption. These phenomena typically occur on time scales ranging from subpicoseconds to a few picoseconds. Once the number of species is determined a kinetic model should be formulated to extract spectra of the species (principal components) contributing to the data. To accomplish this the transients of the kinetic model are fitted to weighted transients S' × V' (the matrix of reduced size corresponding to the number of significant components) as linear combinations with parameters of the kinetic model, such as reaction rates, kept free. The coefficients of the linear combinations obtained in the fit are used to extract spectra corresponding to the assumed kinetics (for mathematical details see Ref. [4]).

We have applied principal component analysis to our measurements of 7HQ dissolved in deuterated water-methanol, for $X_{D_2O}:X_{CD_3OD}$ = a) 0.0:1.0, b) 0.3:0.7, c) 0.5:0.5, d) 0.7:0.3 molar fractions in the solvent mixtures. The data matrices of these four measurements are shown in the corresponding parts of Figure S2 as false colour maps. The experimental scans were taken with piecewise-linear distribution of pump-probe delay points. All four maps show similar set of features: a strong, spectrally broad signal at time "zero" (point #150), a delay-independent bleach signature at 1510 cm$^{-1}$, a feature at 1475 cm$^{-1}$ decaying in magnitude, and two features at 1440 cm$^{-1}$ and 1530 cm$^{-1}$ increasing in strength (only included in maps c and d). Additionally, a spectrally broad decaying feature is visible in the measurements a and b. It appears as a sharp change in colour around point #300 due to the increase in time delay increment in the measurement sequence.

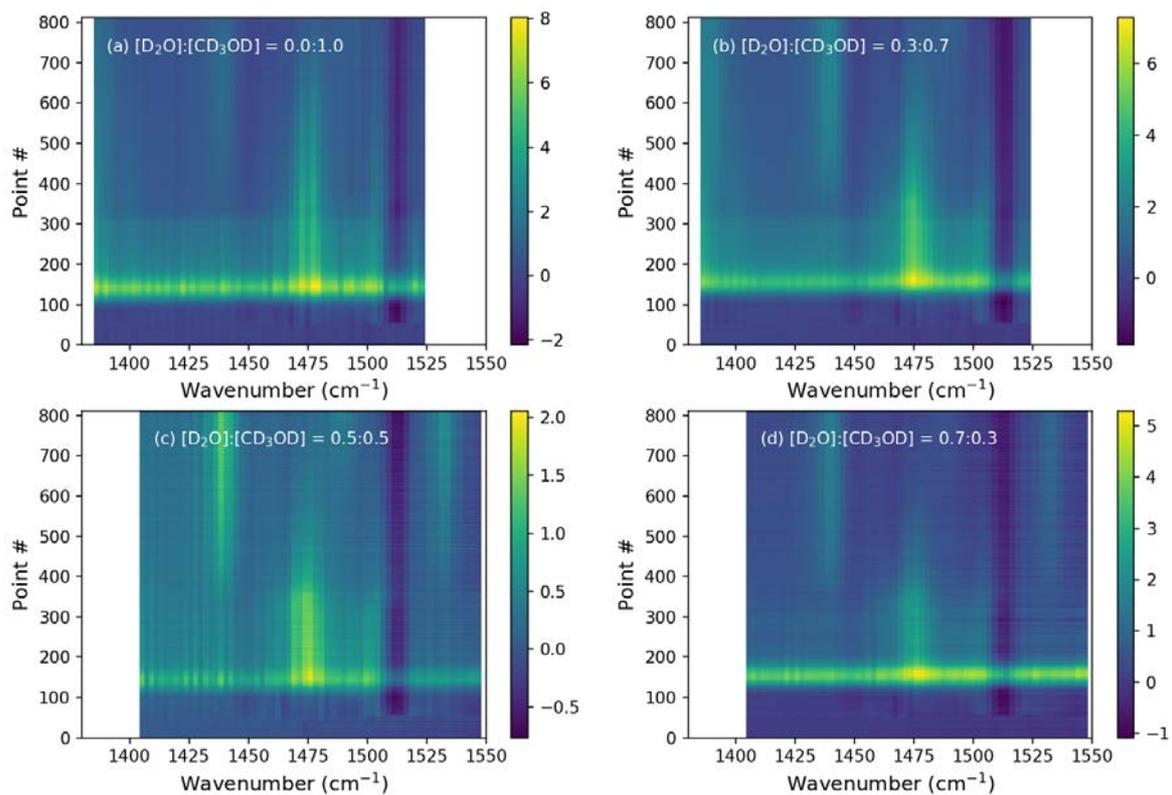

Figure S2: Data matrices presented as false colour maps for the relaxation of 7HQ in the water-methanol solvent mixtures with molar fractions $X_{D_2O}:X_{CD_3OD}$ = 0.0:1.0 (a), 0.3:0.7 (b), 0.5:0.5 (c), 0.7:0.3 (d).

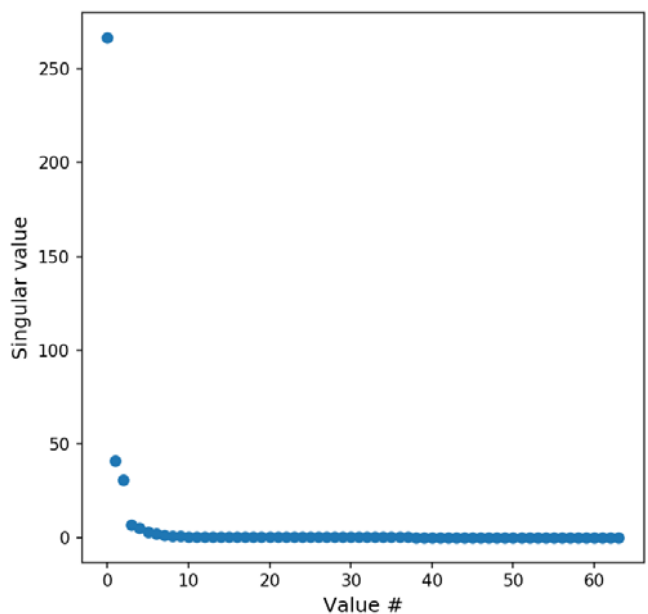

Figure S3: Singular values for the data matrix shown in Figure S1a.

The feature at zero delay corresponds to the solvent response, typically observed in ultrafast UV/IR pump-probe spectroscopy, due to an interplay of cross phase modulation by the nonlinear refractive index and transient absorption features due to free carrier generation by multiphoton absorption in the solvent and flow cell windows. These features peak around "zero" delay and have fully decayed within a time scale of a few picoseconds. In the following we only analyze the data matrices starting from point #200 (delay of about 1 ps) and thus exclude this artifact. The delay-independent bleach component results from contributions of the neutral N form of 7HQ in the electronic ground state.

Figure S3 shows the singular values (diagonal entries of the matrix S) for the measurement in Figure S2a (7HQ in neat $CD_3OD$) in decreasing order. Three first singular values are clearly larger than the rest indicating that the data matrix in Figure S2a contains three significant components. Figure S4 shows these three components as false colour maps in panels (a)-(c). Figure S4d shows the sum of these components. Comparing it with the data matrix reproduced in Figure S4e one can see that the three main components capture all essential dynamics in the recorded transient spectra. Figure S4f shows the residuals. Note that the maximum value on the colour bar constitutes only about 5% of the maximum value in the data matrix. The residuals show two weak components around the wavenumbers 1460 $cm^{-1}$ and 1490 $cm^{-1}$, with spectra shifting towards larger wavenumbers at longer time delays. These features correspond to a spectral diffusion and, as mentioned above, cannot be captured using the principal component analysis. Therefore in the following we neglect these weak features.

As described in Ref. [4] and references therein, a kinetic model needs to be assumed to extract principal components contributing to the recorded data from the components of the SVD analysis. Considering the main three features observed in all data matrices, we assume the following kinetic model: a component with a decay time $\tau_1$, which will represent the spectrally broad fast decay, a component with a decay constant $\tau_2$, corresponding to the main decaying feature at 1475 $cm^{-1}$ and a corresponding rising component. These three transients are described as follows:

$$T1(t) = \exp(-t/\tau_1) \quad , \quad \text{(S2a)}$$
$$T2(t) = \exp(-t/\tau_2) \quad , \quad \text{(S2b)}$$
$$T3(t) = 1-\exp(-t/\tau_2) \quad . \quad \text{(S2c)}$$

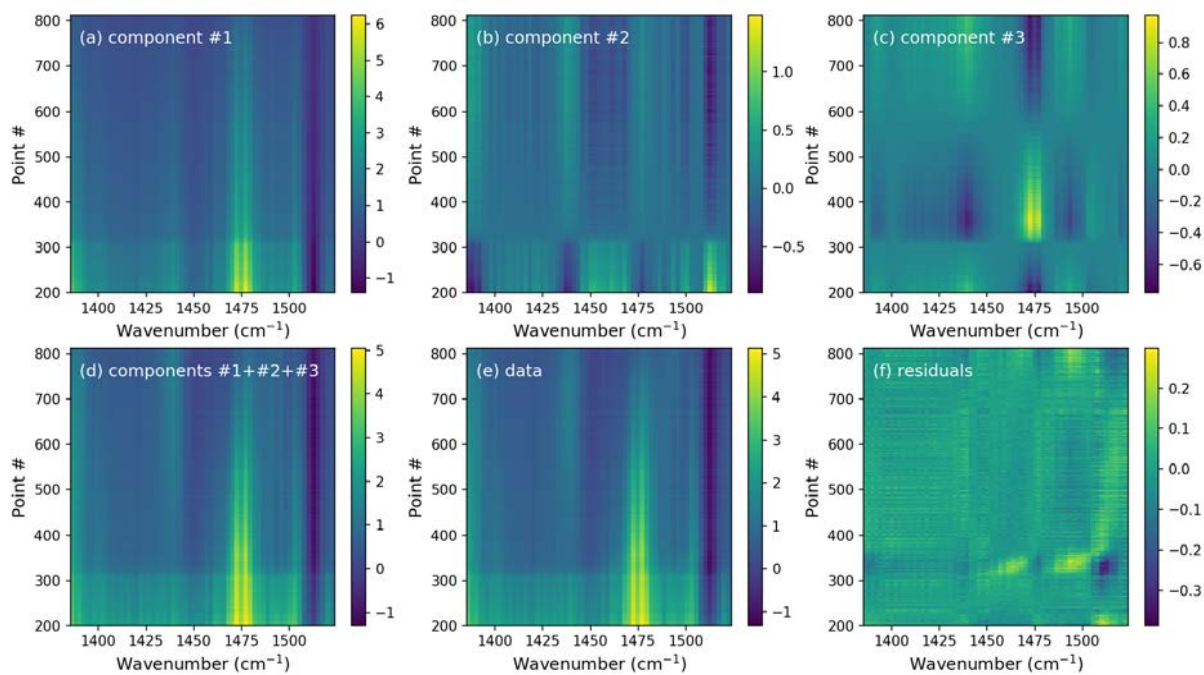

Figure S4: Singular value decomposition of the data matrix 7HQ in neat $CD_3OD$ (depicted in panel (a) of Figure S1): (a) SVD component #1, (b) SVD component #2, (c) SVD component #3, (d) sum of the three components, (e) data matrix of Figure S1a shown here for comparison purposes, (f) residuals upon subtraction of the three components from the data matrix.

Fitting these transients to linear combinations of the three components of the matrix S' × V' the values of the decay constants $\tau_1$ and $\tau_2$ as well as the coefficients of the linear combinations are extracted. The extracted decay constants for all four measurements are given in Table S1.

Table S1: Time constants extracted from the PCA/SVD analysis.

| $X_{D_2O}:X_{CD_3OD}$ | $\tau_1$ (ps) | $\tau_2$ (ps) |
|---|---|---|
| 0.0 : 1.0 | 1.74 ± 0.03 | 361 ± 6 |
| 0.3 : 0.7 | 1.71 ± 0.03 | 180 ± 2 |
| 0.5 : 0.5 | 7.5 ± 0.8 | 158 ± 5 |
| 0.7 : 0.3 | 0.89 ± 0.09 | 110 ± 3 |

The decay time $\tau_1$ = 1.7 ps corresponds to CD$_3$OD solvent response. It becomes insignificant for the measurements c) and d) with large fractions of D$_2$O and therefore decay values obtained for these measurements reflect systematic deviations due to the SVD and fitting procedures (no meaningful third component is extracted for these data sets). The long time constant $\tau_2$, on the contrary, is always well represented both in decaying and in rising features. As evident from the Table S1 the time constant is decreasing for larger concentrations of D$_2$O. The spectra of the slow decay and rise components ($\tau_2$) are plotted in Figure S5. All spectra are normalized to the maximum of the signal at 1475 cm$^{-1}$. As expected, the decay component shows one significant peak at 1475 cm$^{-1}$ and the rise component shows two significant features at 1440 cm$^{-1}$ and 1530 cm$^{-1}$ (only present in the measurements c and d due to scan range limits). The feature at 1440 cm$^{-1}$ comprises only 30% of the decay signal in neat CD$_3$OD, but increases in intensity, when D$_2$O is added. It peaks for a molar fraction of $X_{D_2O}$:$X_{CD_3OD}$ = 0.5:0.5, but this is probably due to noise in this measurement.

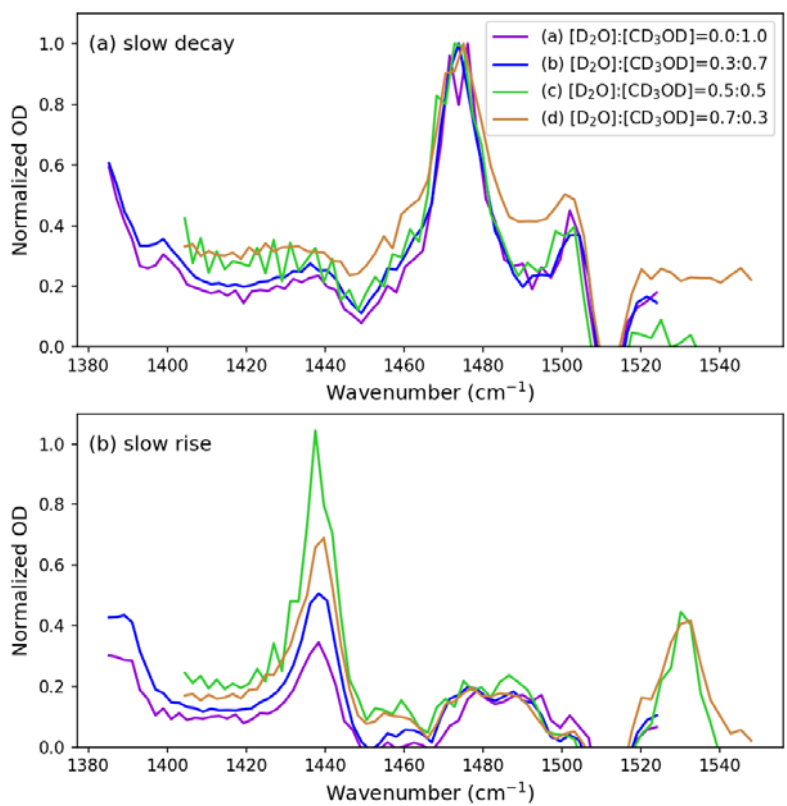

Figure S5: Principal components of the data shown in Figure S1: (a) spectrum corresponding to the slowly decaying component; (b) spectrum corresponding to the slowly rising component.

In concluding this Section, we summarize the findings from the PCA-SVD kinetic analysis of our transient UV/IR pump-probe results obtained on 7HQ in deuterated water-methanol mixtures. Apart from a spectrally broad featureless component with a characteristic decay time constant of 1.7 ps, that we have ascribed to cross phase modulation and multiphoton absorption effects induced in the solvent (and flow cell windows), we observe a bleach signal at 1510 cm$^{-1}$, that we can relate to the N species in the electronic ground state. The positive signal at 1475 cm$^{-1}$, decaying with the solvent composition dependent time constant $\tau_2$ ranging from 361 ps in neat CD$_3$OD to 110 ps for $X_{D_2O}:X_{CD_3OD}$ = 0.7:0.3, is almost equal to the reported value of the 1470 cm$^{-1}$ band for the N* reactant [1]. The positive spectral features located at 1440 cm$^{-1}$ and 1530 cm$^{-1}$, increasing in magnitude with the same time constant $\tau_2$, are indicative of the Z* product species [1]. The direct correlation of the decay kinetics of N* and the rise kinetics of Z*, without any significant contributions from intermediate C* or A* species at intermediate delay times, may suggest that the direct pathway III in Figure 1b prevails. However, as the ensemble averaged N* → Z* deuteron transfer can be described by only one dominating principal component with its kinetic characteristics found in the hundreds of picosecond time scale, much longer than the average lifetimes of hydrogen bonds in water-methanol mixtures (typically on the order of a few picoseconds), we can conclude that a sequential deuteron transfer pathway is the preferred route. We have excluded the protolysis pathway I via the A* intermediate in the previously reported study on 7HQ in neat CD$_3$OD [1], based on the relatively long lifetime for A* (Table 1). The reason why the intermediate C* species, expected to occur for the hydrolysis/methanolysis pathway II, cannot be extracted from the PCA-SVD analysis, likely lies in the fact that the C* → Z* rate is about two times faster than the decay rate of N*, for both neat methanol and neat water (see Table 1) [1, 5, 6]. Hence in a strict sequential kinetic scheme of N* → C* → Z*, the intermediate C* should have a transient population build-up peaking to only about 25% of the initial population of N* excited around a pulse delay time of 0.5·$\tau_2$, and then decaying with about the same time constant $\tau_2$. Hence the signals indicative of the marker bands of C* are not significant enough in magnitude and distinctly different in temporal behaviour to warrant an unequivocal identification.

## 3. Free Energy-Reactivity Analysis

To understand our observations we first discuss the reactivity of acids and bases as a function of solvent medium. Typically acidities in the condensed phase are quantified with p$K_a$-values (i.e. -log$_{10}K_a$, with $K_a$ = [B][H$^+$] / [HB$^+$] , where "H$^+$" is a common indicator of the solvated proton species present in the particular solvent used, and B is the conjugate base of the acid HB), for the water solvent. Empirical data in other solvents (such as methanol, dimethylsulfoxide, acetonitrile, as well as for particular solvent mixtures) now have been gathered, and solvent-dependent acidity relationships have been reported. In particular, a comparison of p$K_a$-values of particular types of acids in water to those in methanol has shown that a strong variation occurs for phenol-type of acids, but a significantly smaller solvent dependence happens for protonated nitrogen bases [7-9]:

$$pK_a \text{ (CH}_3\text{OH)} = m\, pK_a \text{ (H}_2\text{O)} + c \qquad\qquad , \qquad\qquad (S3)$$

where for phenols $m$ = 1.08 and $c$ = 3.66, while for protonated nitrogen bases $m$ = 1.02 and $c$ = 0.72. Hence for phenol-type compounds the p$K_a$-value is observed to increase by ~ 3.5 – 4 units when going from H$_2$O to CH$_3$OH as solvent, whereas for protonated nitrogen bases (amines, anilines, and N-heterocycles such as pyridines) the p$K_a$-value only increases by about 0.5 – 1 unit.

Photoacid research has matured to a general understanding that the photoacidity effect results from an increase of acidity upon electronic excitation, i.e. p$K_a$* = p$K_a$(S$_1$) decreases by ~ 6-7 units compared to the electronic ground state p$K_a$(S$_0$)-value for aromatic alcohols (phenols, naphthols, hydroxypyrenes). A large number of time-resolved fluorescence counting measurements, femtosecond UV-vis and UV-IR pump-probe experiments have led to the conclusion that a free energy-reactivity correlation of photoacids connect the thermodynamic quantities of acidity in the electronic excited state (i.e. p$K_a$*-values) to proton transfer reaction rates. This free energy-reactivity correlation holds for photoacid dissociation to the solvent water, proton abstraction from the solvent water by a photobase, as well as to photoacid-base neutralization in aqueous solution, and has even been shown to hold for proton transfer of photoacids in methanol solution [10-16].

The free energy reactivity correlation can be rationalized in terms of Marcus theory adapted to the case of proton transfer [17], where solvent reorganization plays a dominant role, or by use of the bond-energy bond-order (BEBO) model [18] for proton transfer along a pre-existing hydrogen bond, valid for the opposite extreme condition reminiscent of nonadiabatic electron-transfer reactions. In the free energy ($\Delta pK_a$) parameter value range relevant here, i.e. in the endothermic branch of the proton transfer reactions, these two descriptions lead to similar results, hence we summarize here only the main equations following the BEBO-model. In the free-energy relationship:

$$k_r = k^* \exp(-\Delta G_a / RT) \qquad , \qquad (S4)$$

where $k_r$ is the first-order rate constant and $(k^*)^{-1}$ is the frequency factor of this type of reaction, R is the gas constant and $T$ the absolute temperature. The effective activation energy of the proton transfer reaction, $\Delta G_a$, has been estimated using the Marcus BEBO equation [19]:

$$\Delta G_a = \Delta G^o/2 + \Delta G_o^{\#} + \Delta G_o^{\#} \ln(\cosh[\Delta G^o \ln 2 / (2\Delta G_o^{\#})]) / \ln 2 \qquad . \qquad (S5)$$

Here $\Delta G_o^{\#}$ is the solvent-dependent activation energy of the charge-exchange reaction when the total free energy change $\Delta G^o$:

$$\Delta G^o = RT \cdot \ln 10 \cdot \Delta pK_a \qquad , \qquad (S6)$$

in the proton-transfer reaction is equal to zero.

Free energy-reactivity correlations can be used to derive an unknown $pK_a$-value of one of the acids/conjugate acids of bases in the acid-base neutralization reactions, as has been demonstrated for carbonic acid [13]. In this report we have applied these now well-established free energy-reactivity correlation relationships to conclude that the proton transfer hydrolysis/methanolysis pathway II dominates for 7HQ in water-methanol solvent mixtures. Figure 3 shows the free energy-reactivity correlations for the N*, C* and A* forms of 7HQ in $H_2O$, $CH_3OH$ and $CD_3OD$, as well as for [$D_2O$]:[$CD_3OD$] solvent mixtures. Tables 1 and S1 summarize our findings using our current results in the deuterated water-methanol mixtures, and literature values for 7HQ in either aqueous or methanol solution [1,5,6].

Table S2 shows the resulting values for $\Delta pK_a = pK_a$(donor) − $pK_a$(protonated acceptor) from the time constants derived from the PCA-SVD analysis of the ultrafast UV/IR pump-probe measurements on 7HQ in the deuterated water-methanol mixtures, using the Marcus BEBO free energy-reactivity relationship. For 7HQ in neat $D_2O$ we have derived a value for the $\Delta pK_a$-value from the reported value obtained by Bardez in neat $H_2O$ [6], assuming a regular H/D kinetic isotope effect takes place, i.e. $\tau_{D_2O} = 1.4\ \tau_{H_2O}$.

Table S2  Free-energy reactivity results obtained on 7HQ in deuterated water-methanol.

| $X_{D_2O}$:$X_{CD_3OD}$ | N* decay time constant (ps) | $-\log_{10}[k_r]$ | $\Delta pK_a$ | Pathway I | | | Pathway II | |
|---|---|---|---|---|---|---|---|---|
| | | | | Calculated $pK_a$ (N* species) | $pK_a$ (solvent species $D_3O^+$ or $CD_3OD_2^+$) | $pK_a$ (solvent species $D_2O$ or $CD_3OD$) | Calculated $pK_a$ (C* Species) | |
| 0.0 : 1.0 | 361 | 9.44 | 1.40 | 1.4 | 0.0 | 17.5 | 16.1 | |
| 0.3 : 0.7 | 180 | 9.74 | 1.00 | 1.0 | 0.0 | 15.4 | 14.4 | |
| 0.5 : 0.5 | 158 | 9.80 | 0.86 | 0.9 | 0.0 | 15.0 | 14.1 | |
| 0.7 : 0.3 | 110 | 9.96 | 0.65 | 0.7 | 0.0 | 15.0 | 14.3 | |
| 1.0 : 0.0 | 51 | 10.28 | 0.0 | 0.0 | 0.0 | 15.0 | 15.0 | |

Figure S6 shows the derived effective $\Delta pK_a$-values for the [$D_2O$]:[$CD_3OD$] solvent mixtures as a function of molar fraction of $CD_3OD$. From $\Delta pK_a = pK_a$(donor) - $pK_a$(protonated acceptor), we are able to assess the reactivity of N* as a deuteron acceptor (when $\Delta pK_a = pK_a$($D_2O$ or $CD_3OD$) - $pK_a$(C*) ), or of N* as deuteron donor (when $\Delta pK_a = pK_a$(N*) - $pK_a$($D_3O^+$ or $CD_3OD_2^+$) ), see Table S2. We use reported values for the autoprotolysis constants of $H_2O$, $D_2O$ and $CH_3OH$, as well as for water-methanol mixtures [9,20-22] to specify the acidities of $D_3O^+$ or $CD_3OD_2^+$, and of $D_2O/CD_3OD$ in the deuterated water-methanol mixtures. Here $pK_a(D_3O^+) = pK_a(CD_3OD_2^+) = 0$, i.e. we use the acidity value of $D_3O^+/CD_3OD_2^+$ without taking into account the self-concentration of water or methanol [14,23]. From these reported values we also learn that the acidity of the solvent molecules in the deuterated solvent mixtures remains close to that of neat $D_2O$, down to molar fractions of $D_2O$ as low as 0.2, and only significantly changes – albeit modestly in absolute magnitude – when reaching the neat $CD_3OD$ case. For either proton transfer scenario, as shown in Figure 1, the resulting derived changes for the $pK_a^*$-value for either N* (as deuteron donor in the N* → A* step in Pathway I) or C* (as deuteron acceptor in the N* → C* step in Pathway II) as a function of solvent composition remains modest in magnitude. The overall change of only 1.4 $pK_a$ units when going from $D_2O$ to $CD_3OD$ as a solvent is only compatible with the protonated nitrogen base character of C*, and not with the naphthol acid character of N*. We conclude that the almost linear dependence of $\Delta pK_a$ as a function of molar fraction of $CD_3OD$, with a slope consistent for proton abstraction by the quinoline moiety, reflects a rate determining N* → C* reaction step in the N* → C* → Z* dominating pathway for the whole range of $D_2O$-$CD_3OD$ solvent mixtures.

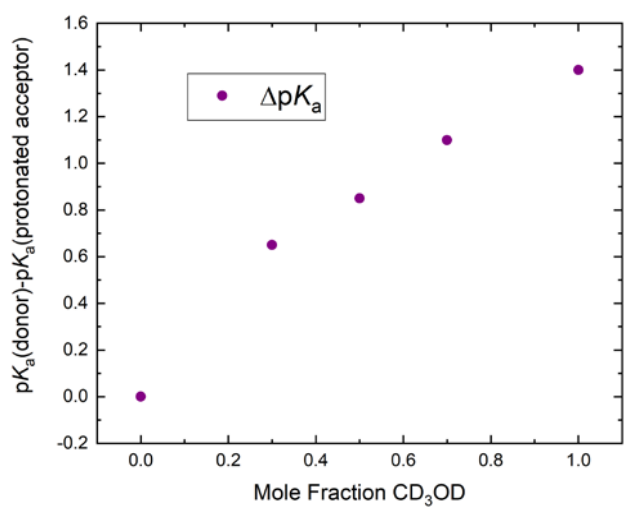

Figure S6: Effective $\Delta pK_a$ values derived from the reaction kinetics of 7HQ in [$D_2O$]:[$CD_3OD$] solvent mixtures, using the free energy reactivity relationship as defined by the Marcus BEBO model as depicted in Figure 3b.

## 4. Computational Details

### 4.1 Equilibration

Molecular dynamics (MD) simulations of 7HQ solvated in methanol-water mixtures were set up at two different mole fractions of methanol, namely 50 % and 30 %. The simulations were staged in a cubic box with side lengths 24.2 Å and 25.7 Å, respectively, hosting one 7HQ molecule and 141 water and 141 methanol molecules in the first, and 288 water and 124 methanol molecules, respectively. In a first step, the simulation boxes were equilibrated in the NPT ensemble ($T$ = 298 K; $p$ = 1 bar) using classical MD simulations, where the TIP3P [24] water force field was used together with a force field for methanol by Fischer *et al*. [25]. For the parametrization of 7HQ, we used the CHARMM-GUI [26] and the CGenFF force field [27]. This procedure resulted in densities of 0.833 g/ml and 0.915 g/ml for 50% and 30 % methanol, respectively, which are close to the respective experimental densities of methanol water mixtures at 25 °C of (~0.88 g/ml and ~0.92 g/ml) [28]. For each mole fraction, we sampled five configurations at evenly spaced intervals from the last 5 ns of the simulations. In a second step, *ab initio* molecular dynamics (AIMD) simulations were run using these 10 configurations as starting points. After around 10 ps of simulation in the NVT ensemble at 350 K using a Nosé-Hoover chain thermostat [29-31], we switched to the NVE ensemble, where the systems were propagated for another 30 ps. For the AIMD simulations, we used the GPW scheme [32] as implemented in the CP2K software package [33]. The BLYP functional [34, 35] with a TZVP valence basis set [36], Goedecker potentials [37], and a 350 Ry plane-wave cutoff was employed. With the increase in temperature we aim to balance overstructuring effects, found in simulations at lower temperatures [38, 39]. For the very same reason, we used DFT-D2 dispersion corrections [40]. The simulation temperature of 350 K corresponds to a physical temperature of 300–320 K. For the ground-state simulations, we used restricted Kohn–Sham density functional theory (DFT) and a time step of 1 fs. All protons in the system were given the mass of deuterium to allow us to use a 1 fs time step.

## 4.2 MD Simulation of 7HQ in the First Excited State

A hybrid linear response time dependent DFT (TD-DFT) molecular mechanics (MM) approach was adopted for carrying out adiabatic MD simulations of ultrafast protonation events of 7HQ in the first excited electronic singlet ($S_1$) state. For this approach the linear response TD-DFT functionality of ORCA 4.0 [41, 42] was combined with the capability of pDynamo for running hybrid QM/MM MD simulations [43]. TD-DFT calculations employed the long-range corrected ωB97X-D3 [44] exchange-correlation functional and the TZVP basis set. To speed up calculations, it was additionally made use of the RIJCOSX [45-47] approximation and the def2/J Coulomb fitting basis set [48]. The number of calculated excited states was restricted to 8.

The starting configurations for the excited state MD simulations were generated from ground state AIMD snapshots that exhibited a three-membered solvent wire between photoacidic and -basic 7HQ moieties. Based on these snapshots, clusters consisting of 7HQ and surrounding solvent molecules were created by cutting a sphere with a radius of 13 Å around the central solvent molecule in the wire, leaving all molecules within the sphere whole. Next, the resulting configurations were modified in two distinct ways by altering the position of a proton to generate (I.) a 7HQ cation with a hydrogen-bonded $HO^-/CH_3O^-$ ion at the nitrogen site, and (II.) a 7HQ anion with an $H_3O^+$ molecule hydrogen bonded to the $RO^-$ site. The `transferred` proton was placed at the equilibrium N-H and O-H equilibrium bond distance, respectively. Three independent simulations were run for each of the two starting points on the basis of three independent ground-state MD snapshots. Note, all protons were given the mass of deuterium to allow for a time step of 1 fs. Nuclear velocities were reassigned to all atoms from the ground-state MD snapshots and the system was propagated on the excited state potential energy surface for 0.5 to 1.5 ps without any coupling to an external heat bath.

The QM part consisted of 7HQ, a three-membered solvent wire connecting the photobasic and photoacidic moieties of 7HQ, and the first solvent layer around the wire. All remaining solvent molecules were treated classically and interact with the QM part only via constant point charges.

## 4.3 MD Simulation of the solvent ion in the presence of 7HQ-C*

To gain mechanistic insights into the charge migration inside the solvent over times of around 10 ps, we carried out additional MD simulations under periodic boundary conditions using a hybrid quantum/classical approach. At this time, the solvent was treated entirely at the DFT level employing revPBE0 [49-51] as exchange-correlation functional together with the D3 dispersion correction [52]. Note, that the importance of using exact Hartree-Fock exchange for modelling the migration of hydroxide ions in aqueous solution was recently pointed out by Chen *et al.*, who found a clear dependence of the solvation shell structure on GGA and hybrid XC functionals [53]. The auxiliary density matrix method [54] (ADMM) was used as implemented in CP2K to allow for efficient calculations of Hartree-Fock exchange, employing the cFIT3 auxiliary basis set. The Coulomb operator [55] was truncated at 6 Å and Kohn-Sham orbitals were expanded in the TZVP basis set and dual-space Goedecker-Tetter-Hutter pseudopotentials [37] were used to represent the atomic cores. A plane wave cutoff of 400 Ry was employed to represent the electron density. 7HQ-C* was described by classical force field potentials (for details on the force field see Section 4.1 Equilibration) and partial charges were obtained from a restrained electrostatic potential fit based on the $S_1$ electronic density of 7HQ. The side lengths of the MM box were equal to the lengths of the respective QM box and full periodic boundary conditions were applied to MM and QM subsystems, respectively. Simulations were started from snapshots obtained from the ground state *ab initio* simulations described above using equal system sizes. To investigate whether the solvent wire plays any role for the directionality of the PT, only initial configurations were selected for which a three-membered solvent wire connecting the reactive groups of 7HQ was formed. In a next step, the position of the solvent proton that was participating in a hydrogen bond to the nitrogen atom of 7HQ was modified analogously to the TD-DFT MD simulations above, i.e. by placing the proton at equilibrium N-H distance at the chromophore, leaving a negatively charged solvent ion ($CH_3O^-$/$OH^-$) hydrogen bonded to the 7HQ nitrogen site. Finally, velocities were reassigned from the previous step and trajectories analyzed after an initial equilibration phase of 1 ps. No thermostat was applied.

The goal of our simulations was to trace the charge migration through the solvent in proximity to the 7HQ chromophore and to elucidate the role of the hydrogen-bonding network of the solvent on the charge migration mechanism. For this, the definition of the asymmetry parameter $\delta$ for proton location along a hydrogen between donating and accepting sites is shown in Figure S7. Here, 7HQ was modeled by classical force field potentials, where we reparametrized the 7HQ partial charges with a restrained electrostatic potential fit, using the $S_1$ electron density as a reference. This treatment implicitly assumes successful acid-base dissociation, i.e. no back-donation to the solvent occurs after the protonation of the nitrogen site. After a short equilibration phase, the charge migration was simulated for 10 ps starting from five different initial conditions.

The MD trajectories reveal that the negative charge, i.e. $CH_3O^-$ or $OH^-$, migrates through the solution with participation of both methanol and water molecules. Table S3 presents the averaged occurrences of all possible partial reactions. Here, a pronounced difference between the two mixing ratios is observed: In the case of the $X_{H_2O}:X_{CH_3OH}$ = 0.5:0.5 solutions, the charge migrates dominantly between methanol or methoxide species, whereas for the $X_{H_2O}:X_{CH_3OH}$ = 0.7:0.3 solutions, charge transfer is almost exclusively observed between water molecules. Therefore, at higher water concentrations the charge migrates preferentially via water molecules whereas methanol species play only a minor role.

Table S3  Average occurrences of distinct transfer reactions in simulations derived for two $X_{H_2O}:X_{CH_3OH}$ molar ratios

| Transfer reaction | Average count for $X_{H_2O}:X_{CH_3OH}$ = 0.5:0.5 | Average count for $X_{H_2O}:X_{CH_3OH}$ = 0.7:0.3 |
|---|---|---|
| $H_2O + OH^- \rightarrow OH^- + H_2O$ | 0.5 | 9 |
| $MeOH + MeO^- \rightarrow MeO^- + MeOH$ | 5 | 0.3 |
| $MeO^- + H_2O \rightarrow MeOH + OH^-$ | 3 | 3 |
| $MeOH + OH^- \rightarrow MeO^- + H_2O$ | 4 | 1.3 |

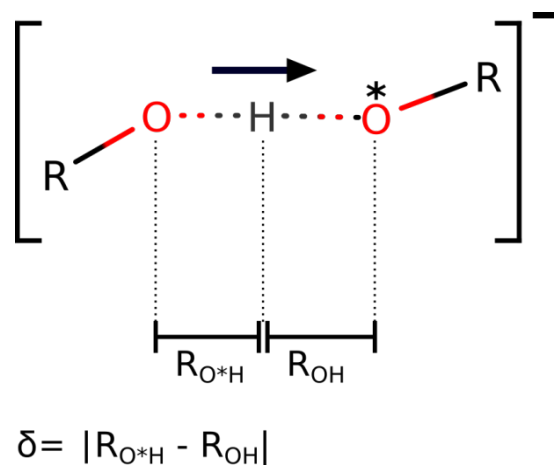

$$\delta = |R_{O^*H} - R_{OH}|$$

Figure S7: Definition of the asymmetry parameter $\delta$ for proton location along a hydrogen between donating and accepting sites, as used to determine the average number of donating hydrogen bonds along the solvent bridge as a function of $\delta$ (see Figure 5, right panel).

## 5. Microsolvation around 7HQ

The local distribution of water and methanol molecules in direct proximity of the 7HQ chromophore differs significantly from the homogeneous distribution. As a consequence, the C* or A* formation, the migration of intermediate solvent ions, and the formation of the N* state might occur at local densities and mixing ratios that are different from the bulk. To analyze the extent of this effect, we computed radial pair distribution functions (RDFs) that inform about radial density variations relative to the bulk from ab initio MD simulations of 7HQ-N in the electronic ground state. The RDFs in Figure S8a indicate a slightly locally increased methanol density for both mixing ratios that extends up to 8-10 Å from the 7HQ center of mass. This effect is more pronounced for the 0.7:0.3 solution, indicating that larger solvation induced changes occur in this case to accommodate 7HQ in the more polar solvent environment. By contrast, partial density variations of water molecules are much less pronounced. The most significant feature in the RDFs of water is seen in the case of the 0.5:0.5 solutions. Here, the RDF between water oxygen atoms and the 7HQ center indicates a slightly increased density, which can be explained by hydrogen bonding of water molecules to the 7HQ hydroxyl moiety.

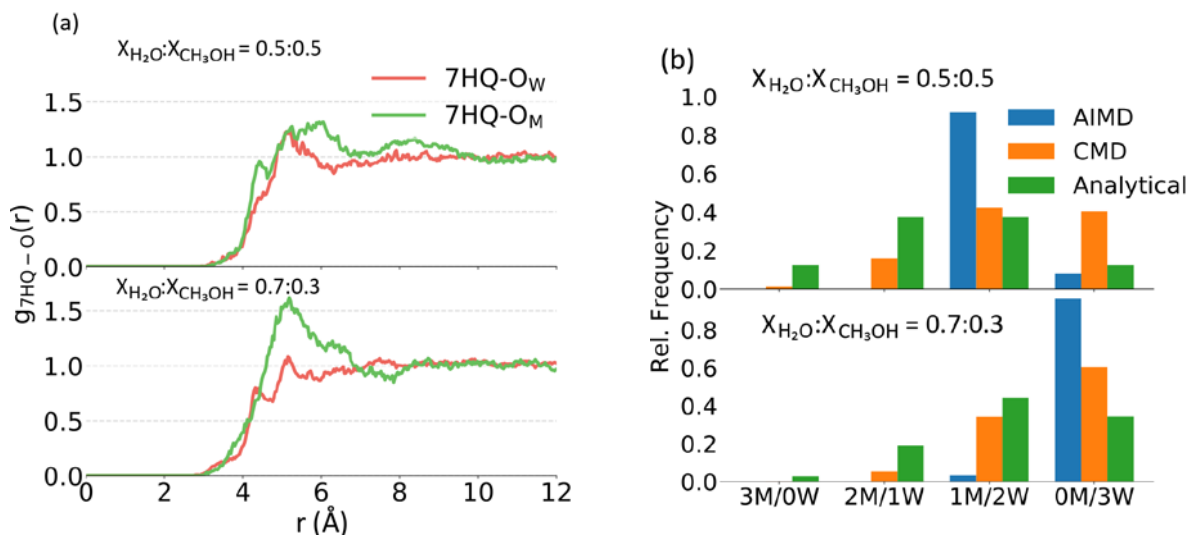

Figure S8: Left: Radial pair distribution functions $g_{7HQ\text{-}O}(r)$ (RDFs) of water ($O_W$) and methanol ($O_M$) oxygen atoms and the center of mass of 7HQ. Right: histograms of the solvent wire composition for BLYP ab initio MD simulations (AIMD), classical MD simulations (CMD), and the analytical distribution arising from counting the various combinations of finding three methanol and zero water molecules (3M/0W) etc. in the solvent wire for the given molar ratios (Analytical).

To conclude the RDF analysis, the solvation of 7HQ alters the spatial distribution of solvent molecules, especially of methanol molecules, however, the effect is rather moderate. This can be seen by comparing the 'local' mixing ratios as calculated from the cumulative number of oxygen atoms up to a radius of 7 Å from 7HQ. The resulting values of 46:54 and 64:36 show little deviation from the overall bulk ratio of 0.50:0.50 and 0.70:0.30 simulations, respectively.

Although this site-unspecific microsolvation is not expected to significantly affect the migration of solvent ions in proximity to the chromophore, 7HQ is able to accept and donate hydrogen bonds at its photoacidic and photobasic moieties, which might alter the solvation structure more site specifically. Most importantly, the well-defined separation of these groups allows for the formation of stable hydrogen bonded chains of solvent molecules. It has been previously shown by us that 'solvent wires' which are comprised of three water molecules are particularly stable over several tens of picoseconds [56]. Hence, these configurations might serve as a potential channel for a sub-picosecond tautomerization reaction. In this case, the system would be in a resting state until a solvent wire forms at which point the tautomerization reaction would take place on a sub-picosecond time scale, involving the concurrent transfer of all protons within the wire [57]. For this reason, information regarding to what extent these wires are already present in the ground state might provide important clues about whether this ultrafast mechanism might play a role at all in the case of more complex solutions than neat water, such as the current methanol water mixtures. To this end, we analyzed *ab initio* and classical MD trajectories of N and investigated the presence of three-membered solvent wires. As shown in Table S4, wires are formed in 7 % of the simulation time in the case of the classical MD simulations, and 9 – 17 % in the case of the AIMD simulations. The discrepancy between the two simulation approaches is likely due to imperfect sampling in the case of the *ab initio* simulations as converged statistics require simulations in the nanosecond regime. Besides the existence of solvent wires, it is compelling to ask whether these configurations are preferably composed of a specific solvent type, i.e., either water or methanol, or whether the distribution of methanol versus water molecules reflects the overall bulk mixing ratio. To answer this, we computed the distribution of water and methanol molecules within the wire, shown in Figure S8b, and compared it to the distribution one would expect by merely counting the various possibilities to form wires of a specific type. As becomes apparent, there is a clear preference for water molecules over methanol molecules to be in the solvent wire than one would expect from plain statistics: Most of the wires consist either of two or three water molecules, whereas wires containing two or three methanol play only a minor role.

Table S4  Average percentages of formation of solvent wires for BLYP *ab initio* (AIMD) and classical MD (CMD) simulations for two $X_{H_2O}:X_{CH_3OH}$ molar ratios.

|  | 0:5:0:5 | 0.7:0.3 |
|---|---|---|
| AIMD | 17 % | 9 % |
| CMD | 7 % | 7 % |

Apart from the solvation structure in vicinity to the chromophore, the charge migration crucially depends on the hydrogen-bonding network between solvent molecules. One obvious difference between water and methanol is that the latter has one hydrogen bonding donor site less. Therefore, the resulting network will exhibit fewer branches the more methanol molecules are present in the solution. Moreover, the higher the number of methanol molecules relative to water molecules becomes, the lower the density will be (cf. ρ(H₂O) = 0.997 g/ml; ρ(MeOH) = 0.791 g/ml at T = 298 K and p=1 atm.).

For a comparison of the density differences for the two mixing ratios, see the RDFs between water and methanol oxygen atoms and all solvent oxygen atoms, respectively, in Figure S9. It can be seen that only the first solvation shell differs, whereas the remaining part of the RDF is almost identical for the two mixing ratios. The number of particles in the first solvation shell, as measured from r = 0 Å until the first minimum, is 3.7 and 4.0 for water and 2.4 and 2.6 for methanol for the 0.5:0.5 and 0.7:0.3 mixtures, respectively. It is clear, that the lower coordination number in the case of the 0.5:0.5 solutions is a consequence of the lower density. Interestingly, in the case of methanol there is a pronounced difference between the ideal coordination number of 3 and the actual coordination number observed in the MD.

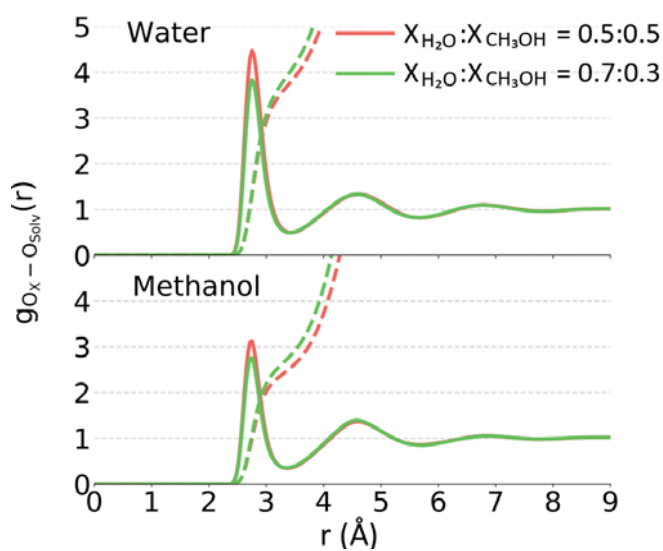

Figure S9: Radial pair distribution functions between oxygen atoms of a specific solvent type (water or methanol) and all the solvent oxygen atoms based on BLYP AIMD simulations in the electronic ground state.